\documentclass[conference]{IEEEtran}
\IEEEoverridecommandlockouts
\usepackage{cite}
\usepackage[normalem]{ulem}
\usepackage{amsmath,amssymb,amsfonts}
\usepackage{algorithmic}
\usepackage{graphicx}
\usepackage{textcomp}
\usepackage{xcolor}
\usepackage{hhline}
\usepackage{titlesec}
\usepackage{multirow}
\usepackage{anyfontsize}
\usepackage{lineno}
\usepackage{subcaption}
\usepackage[hyphens]{url}
\usepackage{amsmath,amssymb,amsfonts}
\usepackage{algorithmic}
\usepackage{graphicx}
\usepackage{textcomp}
\usepackage{xcolor}
\def\BibTeX{{\rm B\kern-.05em{\sc i\kern-.025em b}\kern-.08em
    T\kern-.1667em\lower.7ex\hbox{E}\kern-.125emX}}

\makeatletter

\def\ps@IEEEtitlepagestyle{
  \def\@oddfoot{\mycopyrightnotice}
  \def\@evenfoot{}
}
\def\mycopyrightnotice{
  {\footnotesize
  \begin{minipage}{\textwidth}
  \centering
  Copyright~\copyright~2022 IEEE. Personal use of this material is permitted. However, permission to use this  \\ 
  material for any other purposes must be obtained from the IEEE by sending a request to pubs-permissions@ieee.org.
  \end{minipage}
  }
}    
\begin{document}

\title{Subjective and Objective Quality Assessment of High-Motion Sports Videos at Low-Bitrates}

\author{Joshua P. Ebenezer \thanks{J.P. Ebenezer worked on this project during his internship at Amazon. His current affiliation is with the University of Texas at Austin.}, Yixu Chen, Yongjun Wu, Hai Wei, Sriram Sethuraman\\
Amazon Prime Video, Seattle, WA}

\maketitle

\begin{abstract}
Videos often have to be transmitted and stored at low bitrates due to poor network connectivity during adaptive bitrate streaming. Designing optimal bitrate ladders that would select the perceptually-optimized resolution, frame-rate, and compression level for low-bitrate videos for adaptive streaming across the internet is therefore a task of great interest. Towards that end, we conducted the first large-scale study of medium and low-bitrate videos from live sports for two codecs (Elemental AVC and HEVC) and created the Amazon Prime Video Low-Bitrate Sports (APV LBS) dataset. The study involved 94 participants and 742 videos, with more than 23,000 human opinion scores collected in total. We analyzed the data obtained and we also conducted an extensive evaluation of objective Video Quality Assessment (VQA) algorithms and benchmarked their performance, and make recommendations on bitrate ladder design. We're making the metadata and VQA features available at https://github.com/JoshuaEbenezer/lbmfr-public.
\end{abstract}


\section{Introduction}

 Live video is expected to have reached 13.2\% of global internet traffic by the end of 2021, up from 3.3\% in 2016~\cite{cisco}. Sports' streaming is a subset of live video traffic and is expected to grow from a 18 billion USD market in 2020 to an 87 billion USD market by 2028~\cite{vmr}. Streaming services have to make decisions on what resolution, frame rate, and compression level must be used to stream videos to a customer with a certain available bitrate such that the delivered video is perceptually optimal. Live Sports streaming services such as Amazon Prime Video and NBC Peacock are rapidly growing in areas with limited or developing internet connectivity. Even for users with connections that are high-bandwidth on average, temporary slowdowns in internet speed can occur which may necessitate livestreaming sports' events at lower bitrates. However, perceptual data on low-bitrate high-motion contents is scarce and to our knowledge no subjective video quality database exists that is designed for this use-case.
 \par 
 At higher bitrates and higher resolutions (more than 1080p), studies~\cite{debattista2018frame,ebu} have found that increasing the frame rate is preferred over increasing the resolutions for a fixed bitrate budget. For example, for a wide variety of content, a video streamed at 1080p at 60 fps will be perceptually better than a video streamed at 4K at 30 fps. However, it is unknown whether a transition bitrate-budget point occurs below which higher resolutions are preferred over higher frame rates. In addition, contents with sports or high-motion will need to be treated differently since a higher frame rate might be preferred for a larger range of bitrate-budgets since important events (like a tennis ball being hit) are occurring momentarily in such videos. The gold standard in studying such variations are the perceptual opinion scores provided by humans for different videos of the same content whose parameters such as resolution, frame-rate, and bitrate are varied. We created such a database of 742 videos rated by a total of 94 participants in order to provide an analysis of the low-bitrate regime and also as a resource to train and test objective video quality assessment algorithms.
\section{Related Work}
A number of databases have been created for generic video quality tasks. Crowdsourced databases such as Konvid~\cite{konvid}, YT-UGC~\cite{youtube_ugc}, and VQC~\cite{vqc} are typically used to design video quality metrics for user-generated content. In-lab databases such as LIVE-ETRI~\cite{etri} and LIVE Livestream~\cite{livestream} are created for professionally created content in highly controlled settings and as a result are typically more internally consistent. However, there are very few databases that consider the effect of resolution, compression level, and frame-rate at the same time, and to the best of our knowledge there are no databases specifically targeting the low-bitrate regime for high-motion contents. The LIVE-YT-HFR~\cite{ythfr} database contains 480 videos at 6 frame rates and 5 VP9 constant rate factor levels, but only two resolutions (1080p and 4K). This limits its use for low-bitrate scenarios, where resolutions are typically lower than 720p. The LIVE-ETRI~\cite{etri} database contains 437 HEVC compressed videos spanning 3 frame rates (30, 60, 120 fps), four resolutions (960$\times$540, 120$\times$720, 1920$\times$1080, and 3840$\times$2160) and 5 QP values. Again, the limited number of low resolutions make the database unsuited for low-bitrate ladder design. The AVT-VQDB-UHD1~\cite{avt} database contains 756 videos rated by 104 participants, with resolutions spanning 360p to 4K, framerates from 15 fps to 60 fps, and bitrates from 300 kbps to 15 Mbps. However, the contents in this database are not high-motion sports contents. High-motion contents typically have temporal video artifacts that may not manifest as strongly in low-motion contents. This affects bitrate ladder decisions since the interplay of compression, resolution, and frame-rate is affected by temporal artifacts. Towards filling this important need, we created the first large-scale, in-lab database of high-motion live sports contents encoded at medium and low bitrates. 

\section{Details of Subjective Study}
\subsection{Source Sequences}
\begin{figure}
\begin{minipage}{0.48\linewidth}
  \centering
  \centerline{\includegraphics[width=\linewidth]{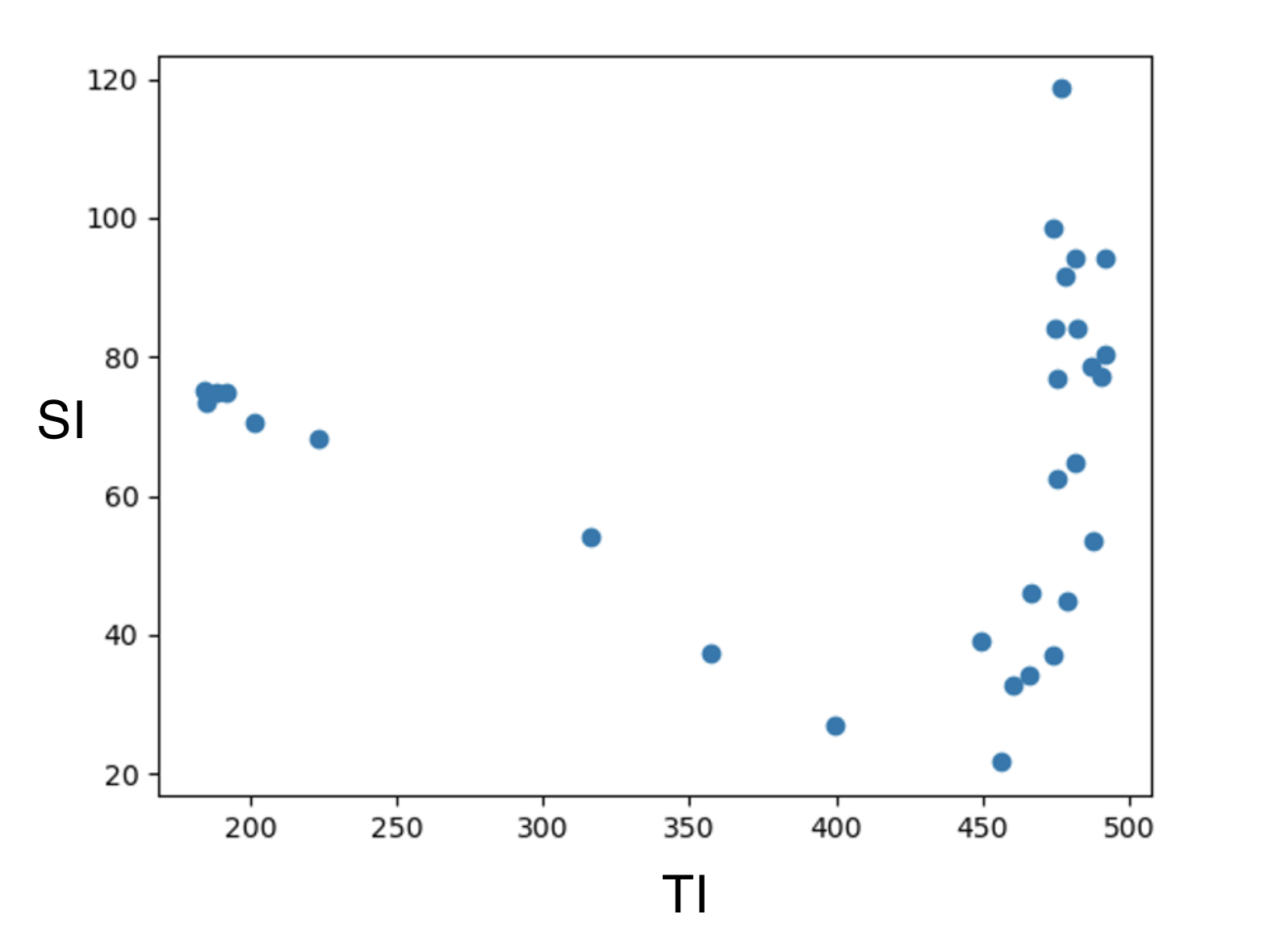}}
 \medskip
\caption{Plot of SI vs TI}
\label{fig:si_vs_ti}
\end{minipage}
\begin{minipage}{0.48\linewidth}
  \centering
  \includegraphics[width=\linewidth]{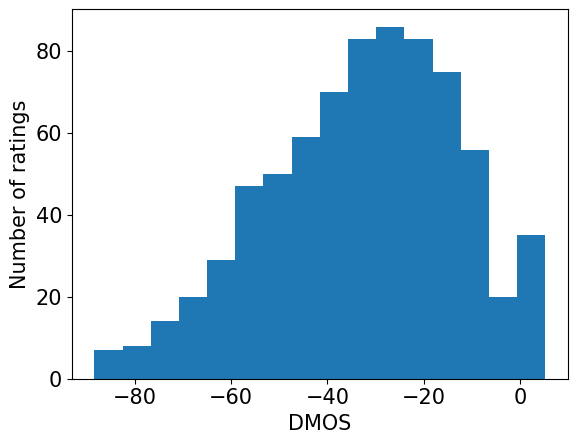}
  \caption{Histogram of DMOS}
  \label{fig:dmos_histogram}
      \end{minipage}
\end{figure}

The 30 source videos were all of sports contents captured from live feeds of sports broadcasting partners. Nine videos were of American football events, eight videos were of soccer, seven videos were of tennis, and six videos were of cricket. The videos covered diverse scenes, including of players running, touchdowns, goals, rallies, slow-motion replays, commentators. The American football source videos had a frame rate of 59.94 frames per second (fps). The soccer, tennis, and cricket videos had a frame rate of 50 fps. Each source video sequence was cut from a longer video. In order to cut the videos accurately, the longer videos were first converted to the y4m format and the frames corresponding to the desired source sequence were cut out. By doing so, we were able to avoid issues related to cutting encoded bitstreams. All the source clips were cut to a length of 8s. The diversity of the 8s source contents can also be captured through the Spatial Information (SI) and the Temporal Information (TI) indices, described in ITU Rec. 910~\cite{rec910}. The SI is plotted against the TI in Fig.~\ref{fig:si_vs_ti}. Note that the values of TI span a large range and are clustered around high values ($>450$), indicating the high-motion nature of the contents. For reference, the highest value of TI for the LIVE-ETRI database was reported to be 250 and the highest TI in the LIVE-YT-HFR database was reported to be 35.

\subsection{Encoding Settings}

The frame rates of the source videos (50/59.94 fps) were considered as High Frame Rates (HFR). Each source video was temporally subsampled by dropping alternate frames to create a Standard Frame Rate (SFR) version of 25 fps or 29.97 fps. Both the HFR and SFR versions were encoded at different bitrates and resolutions using the AVC~\cite{avc} and HEVC~\cite{hevc} codecs with the Elemental hardware encoder~\cite{elemental} in constant bitrate mode. The bitrate-resolution (BR) combinations are listed in Table~\ref{tab:encoding} for each frame rate. The same combinations were used for the AVC and HEVC codecs. Each of the 30 source videos was used to generate 24 distorted videos, and hence 750 videos were used in the study. However, after the study began 8 AVC-encoded SFR videos of soccer were found to have unintended black-borders due to a bug in the encoder software, and hence had to be discarded for the final analysis. 6 AVC encoded HFR videos of soccer were also found to have the black-border, but were corrected in time for 24 subjects in group 2 to watch the correct version, and were hence retained for the final analysis with 24 scores.

\begin{table}
\caption{ Bitrate-resolution-framerate combinations for encoding}
\begin{center}
\setlength\tabcolsep{1.5pt}
\begin{tabular}{|c|c|c|c|}
\hline
HFR & SFR \\
\hline
 288p at 300 kbps & 288p at 200 kbps  \\
 396p at 600 kbps &  396p at 450 kbps \\
 396p at 900 kbps &  396p at 800 kbps \\
 540p at 1300 kbps &  540p at 1200 kbps  \\
 540p at 1900 kbps &  540p at 1800 kbps\\
720p at 2200 kbps & 720p at 2000 kbps  \\
\hline
\end{tabular}
\label{tab:encoding}
\end{center}
\vspace{-5mm} 
\end{table}

\subsection{Subjective Study Settings}

The study was conducted at Amazon Prime Video's lab at Seattle over a duration of 4 weeks. Two TVs were placed more than 15 feet apart such that subjects could watch the videos sitting back-to-back without interference. Subjects were seated at a distance of 1.5H from the TV, where H was the height of the TV. The TVs were identically calibrated 55 inch 4K LG C9 OLED TVs which are capable of playing 4K videos at 60 fps. They were attached through HDMI 2.1 as external displays to two Alienware M15 R4 Gaming Laptops with NVIDIA GeForce RTX 3070 GPUs. The windows of the lab were covered with black paper to block outside light and the lights were dimmed to create a low-light setting compliant with ITU Rec 500~\cite{bt500}. 94 participants volunteered to take part in the study. The 750 videos were divided into 3 groups of 250 videos each such that videos of the same content appeared in the same group. The contents were divided across the 3 groups in such a way that each group had an equal representation of each sport. The participants were divided into 3 groups of 32, 30, and 32 participants each. Each group of participants watched a different group of 250 videos. The study was divided into two 30 minute sessions for each participant to minimize fatigue, and participants were allowed to choose when to schedule their sessions. The age range of the participants was 19-40, all participants had normal or corrected-to-normal vision. The first session began with a training session where subjects were shown 3 videos of poor, medium, and good quality, and told about the quality of the videos, with the intention of ensuring that the score distribution would be reasonable. The scores from the training session were discarded. After that, the study would begin and scores for each video would be collected according to the ACR-HR protocol described in ITU Rec. 910~\cite{rec910}. At the end of each video, a rating scale from 0-100 with increments of 1 was presented with 5 equally spaced verbal markers of Poor, Bad, Fair, Good, and Excellent. Subjects could choose any value on the scale using a mouse.

\section{Analysis of Scores}

\subsection{Internal Consistency}
For each video, all the subjects who watched that video were divided into two equal-sized groups randomly, and Z-score for that video was computed for each group separately. This was done for all the videos, generating two sets of Z-scores. The Spearman Rank Ordered Correlation Coefficient (SRCC) was computed between the two sets. This procedure was repeated 100 times, each time doing a different random split of all the subjects who watched a video. The median SRCC of the 100 randomized splits was 0.9543, indicating a high degree of internal consistency. 

\subsection{Recovery of Quality Scores}

\begin{figure*}
\begin{subfigure}{.24\textwidth}
  \centering
  \includegraphics[width=\linewidth]{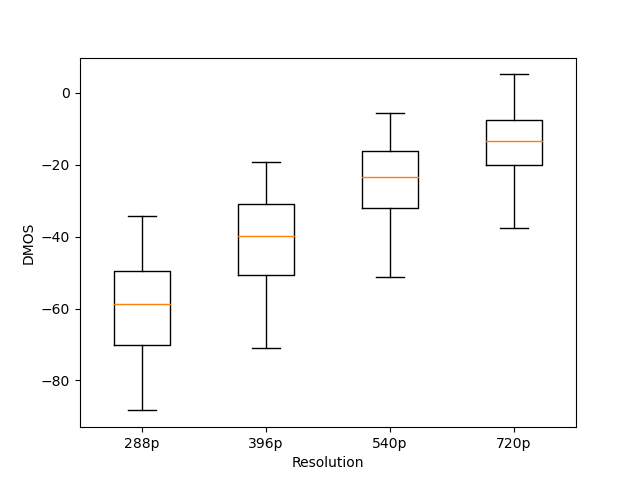}
  \caption{DMOS by Resolution}
  \label{fig:dmos_resolution}
\end{subfigure}%
\begin{subfigure}{.24\textwidth}
  \centering
  \includegraphics[width=\linewidth]{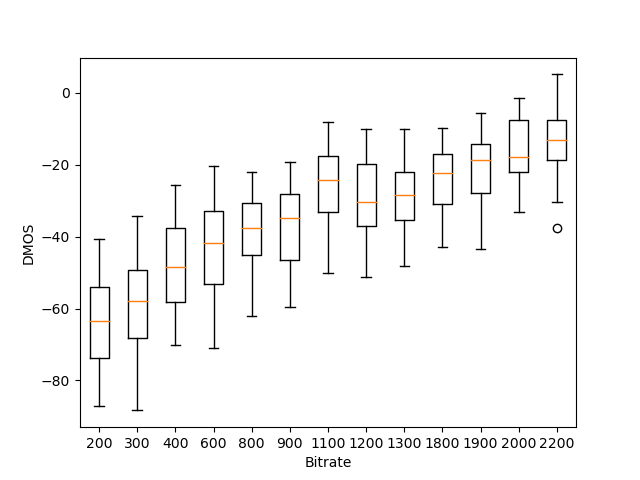}
  \caption{DMOS by Bitrate}
  \label{fig:dmos_bitrate}
\end{subfigure}%
\begin{subfigure}{.24\textwidth}
  \centering
  \includegraphics[width=\linewidth]{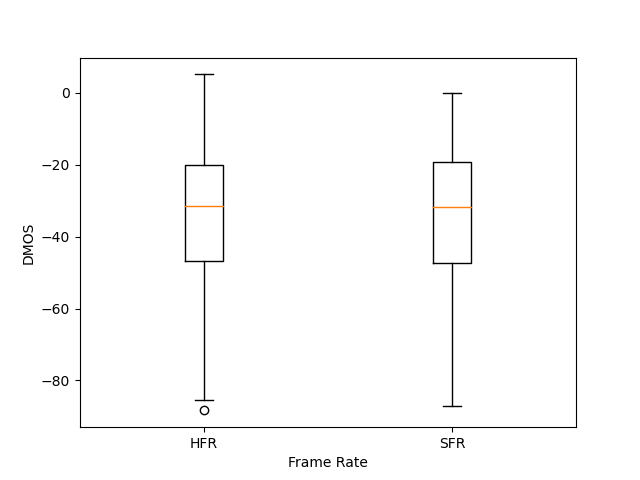}
  \caption{DMOS by Frame Rate}
  \label{fig:dmos_framerate}
\end{subfigure}%
\begin{subfigure}{.24\textwidth}
  \centering
  \includegraphics[width=\linewidth]{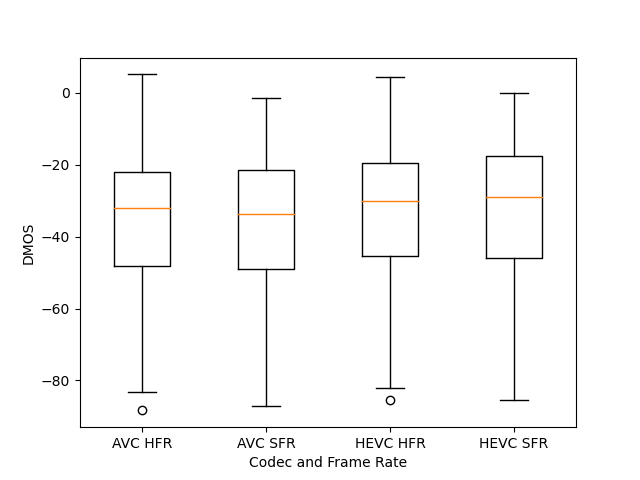}
  \caption{DMOS by Codec and Frame Rate}
  \label{fig:dmos_codec_framerate}
\end{subfigure}
\caption{DMOS by different categories}
\label{fig:dmos_categories}
\end{figure*}

We used the Sureal~\cite{sureal} method to estimate the quality scores from the study using maximum-likelihood estimation. Each opinion score $U_{ij}$ is modelled as 
\begin{equation}
    U_{ij} = \psi_j + \Delta_i + \nu_i X
\end{equation}
 where $\psi_j$ is the true quality score of video $j$, $\Delta_i$ represents the bias of subject $i$, the non-negative term $\nu_i$ represents the inconsistency of subject $i$, and $X \sim N(0,1)$ are i.i.d. Gaussian random variables. The true score, subject bias, and subject inconsistency are estimated through maximum-likelihood estimation with a Newton-Raphson solver. $\psi$ was treated as the Mean Opinion Score (MOS) and the Differential MOS (DMOS) was computed as the difference between the MOS of the distorted video and the MOS of the pristine video. A histogram of DMOS from the study is shown in Fig.~\ref{fig:dmos_histogram} and shows that the distribution is broad. We also plot DMOS according to different categories in Fig.~\ref{fig:dmos_categories}. The scores in each category have a high degree of overlap because of the interplay of the other factors, though the broadly monotonic relationships between resolution and DMOS and bitrate and DMOS are expected. HEVC encoded videos have a slightly higher median DMOS than AVC encoded videos, but with a high degree of overlap, as can be seen from Fig.~\ref{fig:dmos_codec_framerate}.

\par
\begin{figure*}
\begin{subfigure}{.24\textwidth}
  \centering
  \includegraphics[width=\linewidth]{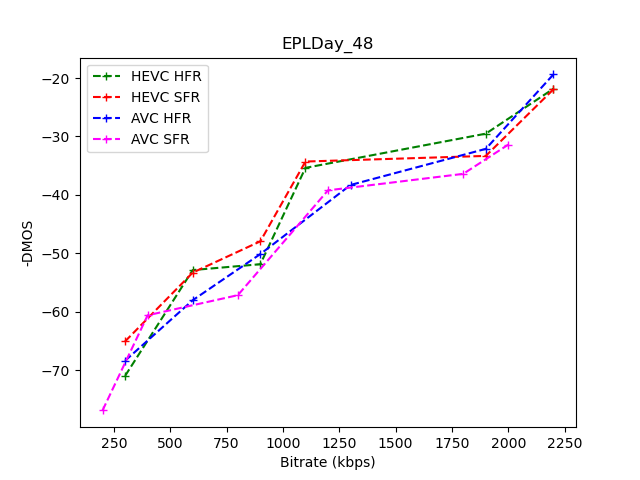}
  \caption{Soccer}
  \label{fig:sfig1}
\end{subfigure}%
\begin{subfigure}{.24\textwidth}
  \centering
  \includegraphics[width=\linewidth]{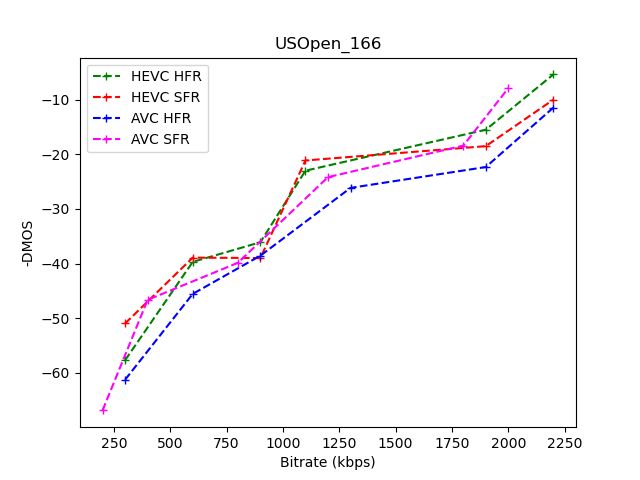}
  \caption{Tennis}
  \label{fig:sfig2}
\end{subfigure}%
\begin{subfigure}{.24\textwidth}
  \centering
  \includegraphics[width=\linewidth]{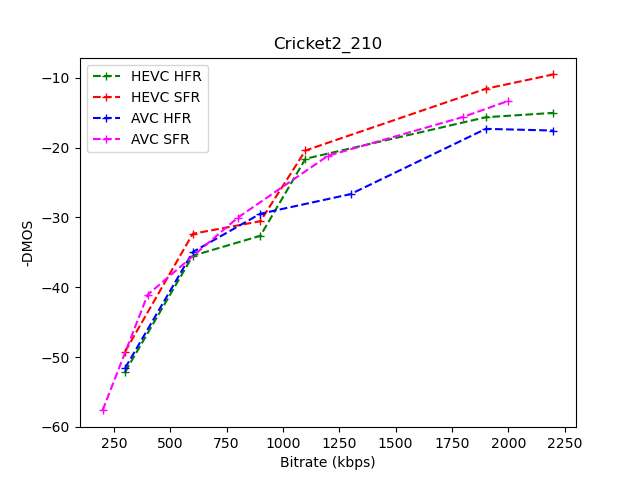}
  \caption{Cricket}
  \label{fig:sfig3}
\end{subfigure}%
\begin{subfigure}{.24\textwidth}
  \centering
  \includegraphics[width=\linewidth]{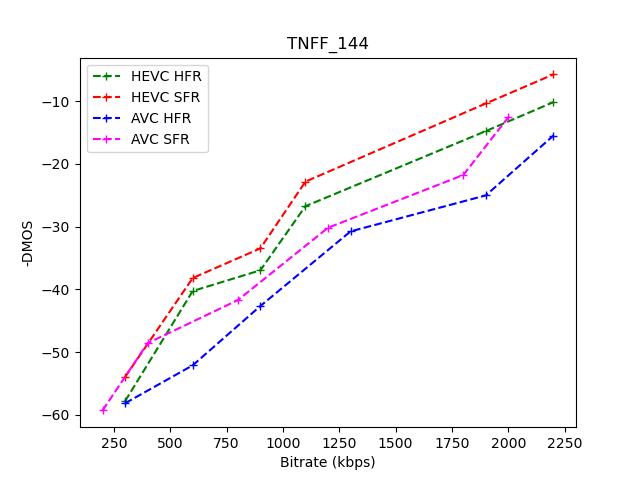}
  \caption{Football}
  \label{fig:sfig4}
\end{subfigure}
\caption{DMOS vs bitrate for 4 contents}
\label{fig:dmos}
\end{figure*}

A few DMOS vs bitrate plots are shown in Fig.~\ref{fig:dmos} for 4 of the 30 contents, with each sport being represented. Each figure is therefore plotted using data for 25 videos and represents quality scores for all the BR combinations for HEVC and AVC codecs for both HFR and SFR. The BR combinations are from Table~\ref{tab:encoding}. Each plot is quite different and this is indicative of the highly content-dependent nature of perceptual quality and spatio-temporal artifacts. A number of important observations can be made from each plot for each content. When a video's frame rate is doubled at the same bitrate while keeping the resolution the same, spatial artifacts increase because the codec has to encode double the number of frames with the same bitrate budget. If the deleterious effect of these spatial artifacts is greater than that of the temporal artifacts caused by high-motion contents played at SFR, the perceptual quality of the SFR version will be greater. For the soccer video encoded with HEVC, HFR is preferred over SFR only after the 540p at 1200 kbps point, indicating that there are indeed such transition points in the low-bitrate regime below which SFR is preferred. On the other hand, for the AVC-compressed tennis video and the HEVC and AVC compressed football videos, the SFR version is always rated better than the HFR version, indicating that the transition point may only appear beyond the 720p at 2000 kbps point. The HEVC-encoded version of the tennis video, however, does have a transition point at 540p at 1300 kbps beyond which HFR is preferred over SFR. The content-dependent nature of the ratings indicates the need for per-title or per-shot bitrate-ladder decisions.

\section{Results on Objective Quality Assessment}

\subsection{Metrics}
 We report the SRCC between the raw quality predictions made by the objective VQA algorithms and the DMOS. We also transform the predictions to the range of the DMOS by using a logistic function
\begin{equation}
f(s) = \beta_1(\frac{1}{2} - \frac{1}{(1+\exp(\beta_2(s-\beta_3))})+\beta_4 x+\beta_5.
\end{equation}
The parameters are found by fitting $f(s)$ to the DMOS. We then compute Pearson's Linear Correlation Coefficient (LCC) and the Root Mean Square Error (RMSE) between the transformed predictions and the DMOS.

\subsection{Full-Reference Algorithms}
\label{sec:fr}
Full-Reference (FR) VQA algorithms compare the distorted video to the pristine video and predict the difference in quality.  We tested a number of FR algorithms with our dataset. All the FR algorithms expected the input distorted and pristine videos to be of the same dimensions. Since the distorted videos in our dataset were not all the same resolution and frame rate as the pristine videos, they had to be spatially and temporally upsampled to match the resolution and frame rate of the pristine videos. Spatial upscaling was performed with fast bilinear upscaling. Temporal upscaling was performed with both frame duplication and motion interpolation, and we found that FR algorithms had an average increase in SRCC of 12\% when their inputs were the motion-interpolated distorted videos when compared to when the inputs were frame duplicated videos.  In Table~\ref{tab:frmint}, we report results obtained for FR algorithms that are either pre-trained or that do not need to be trained for motion-interpolated distorted videos.
\par 
We trained and evaluated ST-GREED~\cite{stgreed}, an FR algorithm that generates 4 features that need to be trained by a Support Vector Regressor (SVR). We also trained and evaluated VMAF from scratch with the motion-interpolated videos. The training procedure was to divide the videos into a training set and a test set with an 80:20 ratio. 5-fold cross-validation was performed on the training set with a grid search to find the best hyperparameters for the SVR. The resulting SVR was evaluated on the test set. This procedure was repeated 100 times with different train-validation-test splits. The median metrics and their standard deviations are are presented in Table~\ref{tab:frtrained}.
\subsection{No-Reference Algorithms}
We evaluate a completely blind No-Reference (NR) algorithm, NIQE, and also train and evaluate ChipQA~\cite{chipqa}, a leading NR VQA algorithm. The results are presented in Table~\ref{tab:nr}. We also evaluate two algorithms that rely on additional metadata: ITU P1204.3~\cite{p1204}, a no-reference algorithm that requires the bitstream information of the distorted video in addition to the video itself, and another feature set that just contains the bitrate, frame-rate, resolution, and codec, which we simply refer to as "Vid. Level metadata". The training procedure described in Section~\ref{sec:fr} was used, with the exception that the algorithms were trained to predict the MOS and not the DMOS, and the results are presented in Table~\ref{tab:nr}. The metadata based algorithms perform well though such metadata is not always available. NR algorithms are especially relevant for livestreaming since the video obtained at the source may not be free of artifacts, as discussed in detail in \cite{livestream}.    

\begin{table}
\caption{SRCC, LCC, and RMSE for untrained/pre-trained FR}
\begin{center}
\begin{tabular}{|c|c|c|c|}
\hline
\textsc{Method}  &  SRCC & LCC & RMSE\\
\hline
PSNR & 0.3544 & 0.3119 & 17.8795 \\
\hline
SSIM\cite{ssim} & 0.4953 & 0.3750 & 17.4286 \\
\hline
MS-SSIM\cite{msssim}  & 0.3416 & 0.0357 & 19.6043 \\
\hline
SPEED\cite{speed} &  0.5749 & 0.5904 & 15.1737\\
\hline
V-SPEED\cite{speed} & 0.3904 & 0.0996 & 18.7071\\ 
\hline
Spatial VIF~\cite{vif} & 0.4581 & 0.4735 & 16.5590 \\
\hline
DLM~\cite{dlm} & 0.8430 & 0.8275 & 10.5541 \\
\hline
VMAF~\cite{vmaf} (pre-trained) & 0.8518 & 0.8325 & 10.4156 \\
\hline 
\end{tabular}
\label{tab:frmint}
\end{center}
\end{table}

\begin{table}
\caption{SRCC, LCC, and RMSE for trained FR}
\begin{center}
\begin{tabular}{|c|c|c|c|}
\hline
\textsc{Method}  &  SRCC & LCC & RMSE\\
\hline
ST-GREED~\cite{stgreed}  & 0.57$\pm$0.08 & 0.59$\pm$0.10 & 14.56$\pm$1.84 \\
\hline
VMAF (trained) \cite{vmaf} & 0.84$\pm$0.10 & 0.84$\pm$0.09 & 9.63$\pm$2.19 \\
\hline
\end{tabular}
\label{tab:frtrained}
\end{center}
\vspace{-3mm} 
\end{table}

\begin{table}
\caption{SRCC, LCC, and RMSE for blind NR VQA}
\begin{center}
\begin{tabular}{|c|c|c|c|}
\hline
\textsc{Method}  &  SRCC & LCC & RMSE\\
\hline
NIQE~\cite{niqe} & 0.4195 & 0.4592 & 15.7479 \\
\hline
ChipQA~\cite{chipqa}  & 0.83$\pm$0.09 & 0.84$\pm$0.09 & 9.31$\pm$2.40 \\
\hline
\end{tabular}
\label{tab:niqe}
\end{center}
\end{table}

\begin{table}
\caption{SRCC, LCC, and RMSE for trained metadata-based NR VQA}
\begin{center}
\begin{tabular}{|c|c|c|c|}
\hline
\textsc{Method}  &  SRCC & LCC & RMSE\\
\hline
P 1204.3 \cite{p1204} & 0.93$\pm$0.14 & 0.93$\pm$0.01 & 6.18$\pm$0.75 \\
\hline
Vid. Level Metadata  & 0.94$\pm$0.02 & 0.94$\pm$0.02 & 6.10$\pm$0.62 \\
\hline
\end{tabular}
\label{tab:nr}
\end{center}
\vspace{-3mm}
\end{table}

\section{Dicussion and Conclusion}

It appears from the data that there isn't a uniform recommendation suitable for all high-motion contents with different properties. Each content has a different concave envelope that represents the best BR combination and frame rate that has to be used for perceptually-optimized viewing. This further underscores the need for VQA algorithms that can predict the quality of videos encoded at different settings and make a decision on the perceptually optimal setting for each content or shot separately. For livestreamed sports, making such content-dependent decisions on the fly, ideally with an automated system, requires FR or NR algorithms with very low latencies. VMAF performs the best among the FR algorithms we tested. Among NR algorithms, both ChipQA and P1204.3 perform very well. P1204.3 performs better than ChipQA, but the difference in their performance is within the standard deviations of their results. If the video is re-encoded in some way at any stage after encoding P1204.3 will fail, whereas ChipQA is a pixel-based method that can work without bitstream information. We also make the observation that comparing  motion-interpolated SFR distorted videos to HFR pristine videos allows FR algorithms to make better judgments than using frame duplication. We are releasing the metadata, VQA features, and scores for our study for researchers to study the effect of different encoding and frame-rate parameters on quality and envision that this database will be an important resource for the community.

{\small
\bibliographystyle{IEEEtran}
\bibliography{conference_101719}
}

\end{document}